\documentclass{wlscirep}

\usepackage{color, graphicx}
\usepackage[T1]{fontenc}
\usepackage[utf8]{inputenc}
\usepackage{etoolbox}

\title{Universal dynamics of mitochondrial networks: a finite-size scaling analysis}

\author[1,*]{Nahuel Zamponi}
\author[2]{Emiliano Zamponi}
\author[3,6]{Sergio A. Cannas}
\author[4,5,6]{Dante R. Chialvo}

\affil[1]{Division of Hematology and Medical Oncology, Department of Medicine, Weill Cornell Medicine. 1300 York Avenue, New York, NY 10065, USA.}
\affil[2]{Department of Molecular, Cellular, and Developmental Biology, University of Colorado-Boulder, Boulder, CO 80309, USA}
\affil[3]{Facultad de Matem\'atica Astronom\'ia F\'isica y Computaci\'on, Universidad Nacional de C\'ordoba, Instituto de F\'isica Enrique Gaviola (IFEG-CONICET), Ciudad Universitaria. (5000) C\'ordoba, Argentina.}

\affil[4]{Center for Complex Systems and Brain Sciences (CEMSC${^3}$), Escuela de Ciencia y Tecnolog\'ia, Universidad Nacional de Gral. San Mart\'in, Campus Miguelete, 25 de Mayo y Francia (1650), San Mart\'in, Buenos Aires, Argentina.}

\affil[5]{Instituto de Ciencias F\'isicas (ICIFI), Escuela de Ciencia y Tecnolog\'ia, Universidad Nacional de Gral. San Mart\'in, Campus Miguelete, 25 de Mayo y Francia (1650), San Mart\'in, Buenos Aires, Argentina.}

\affil[6]{Consejo Nacional de Investigaciones Cient\'ificas y T\'ecnicas (CONICET), Godoy Cruz 2290, (1425) Buenos Aires, Argentina.}
\affil[*]{zamponi.n@gmail.com}

\begin{abstract}
A growing body of evidence suggests that the structure of mitochondrial networks is poised near criticality, an intermediate regime lying in between order and disorder. Such description fits well with the idea that biological systems, in general, may benefit from the long-range correlations and large flexibility conferred by a critical regime. Despite the attractiveness of this proposal, a clear understanding of the possible scenarios leading these networks to criticality is still lacking. In this work, we compared the behavior of mitochondrial networks emerging from a dimensionless agent-based (AB) model and a spatially explicit (SE) model, in which nodes are embedded on a 2D lattice. In both scenarios, we described the position of the control parameter at which mitochondrial networks exhibit a dynamical phase transition as well as the size-dependency of several network features. Furthermore, we showed that the mitochondrial networks from mouse embryonic fibroblasts presented similar topologies to the ones generated using the AB model, while their universal behavior is better described by a SE model. Using finite-size scaling analysis conducted on models and empirical data we defined the universality classes they belong and provided the theoretical boundaries for the mechanisms governing mitochondrial network formation. Our findings predict the full repertoire of dynamical behavior expected for real mitochondrial networks under physiological and pathological conditions.
\end{abstract}

\begin{document}
\flushbottom
\maketitle


\section{Introduction}
The arise of mitochondria constitutes a milestone in the evolution of eukaryotes. Their incorporation into the proto--eukaryotic cell made possible a major increase in genome complexity by allowing the cell to afford the energetic cost of a bigger proteome\cite{lane}. Millions of years of evolution have placed mitochondria not only at the center of energetic and biosynthetic metabolic pathways, but also as essential regulators of homeostasis and cell death\cite{spinelli, sena, eisner, wang}. During that time, most of the mitochondrial genome has been transferred to the nuclear genome, favoring the emergence of complex regulatory networks that constantly match mitochondrial activity with the metabolic demands of the cell while maintaining the organelle's autonomy\cite{timmis, asin-cayuela, couvillion}.

In animals and fungi, mitochondria organizes as a dynamic tubular-reticular network that extends throughout the whole cellular volume. Such network is composed of clusters of different sizes subjected to a constant process of fission and fusion that favors both the maintenance of the stoichiometric relations between electron transport chain (ETC) complexes and a homogeneous distribution of the mitochondrial DNA (mtDNA)\cite{liu, yao, nunnari, aryaman}. Moreover, the dynamic nature of the organelle allows for additional levels of regulation that contribute to mitochondrial homeostasis. For example, a damaged mitochondrial fragment can be either fused to healthy mitochondria and be rescued by content mixing or, if the damage is irreversible, excised from the rest of the network and recycled in a process called mitophagy\cite{twig, okamoto}.

It is well established that mitochondrial morphology is a continuously evolving state arising from constant fission/fusion dynamics, controlled at the molecular level by nuclear encoded proteins whose abundances and activities fluctuate in response to internal and external stimuli\cite{friedman, viana, sabouny}. Despite being extremely dynamic, the overall topology of the mitochondrial ensemble, determined by the distribution of cluster sizes, remains constant, suggesting the presence of an organizing principle at the global scale.

The apparent contradiction between structural robustness and functional susceptibility immediately traces back to critical phenomena, since physical systems poised at the vicinity of a phase transition are know to display such characteristics\cite{bak, mora}, raising the question of whether such competing demands could be mechanistically fulfilled by mitochondria by being close to criticality. The role that critical phenomena may be playing in mitochondrial dynamics have been emphasized already in a number of results, including the phase transition in mitochondrial depolarization as a function of the abundance of reactive oxygen species\cite{aon1}, the percolation-like phase transition in the structure of the mitochondrial condriome as a function of the fission and fusion rates\cite{sukhorukov}, the power-law relation in the mass distribution of mitochondrial clusters\cite{zamponi} and the fluctuation-driven critical tuning of vascular smooth muscle cells mitochondrial networks\cite{bartolak-suki}. However, to demonstrate that mitochondrial function is the result of a certain kind of critical behavior at the cellular level, one should be able either to quantify the changes in the system correlation properties while varying some control parameter or to establish the lack of a characteristic scale by varying the system size. 

In this work we use finite-size scaling analysis to determine the nature of the transition in mitochondrial network dynamics. We first study the universal properties of two classes of models of mitochondrial dynamics with the aim of finding the boundaries for the universal dynamics in real mitochondria. We then use a finite-size scaling strategy to extract the critical exponents from real mitochondrial networks and compared them with the results from our models. By doing this, we demonstrate that mitochondrial dynamics are critical and that they belong to the standard percolation universality class.

The paper is organized as follows. In the next section, the two models used in the study are introduced and the quantities that will be used to characterize them are defined. The results section contains first  a detailed analysis of each of the models including the behavior of the relevant quantities as a function of the control parameter and the finite-size scaling analysis. Similar description is done for the data from real mitochondrial networks extracted from microscopy images. The paper close with a brief discussion on the relevance of the present results to understand mitochondrial dynamics. Further details for the methods are described in a dedicated section.


\begin{figure} [h!]
\centering
\includegraphics[width=.95\textwidth]{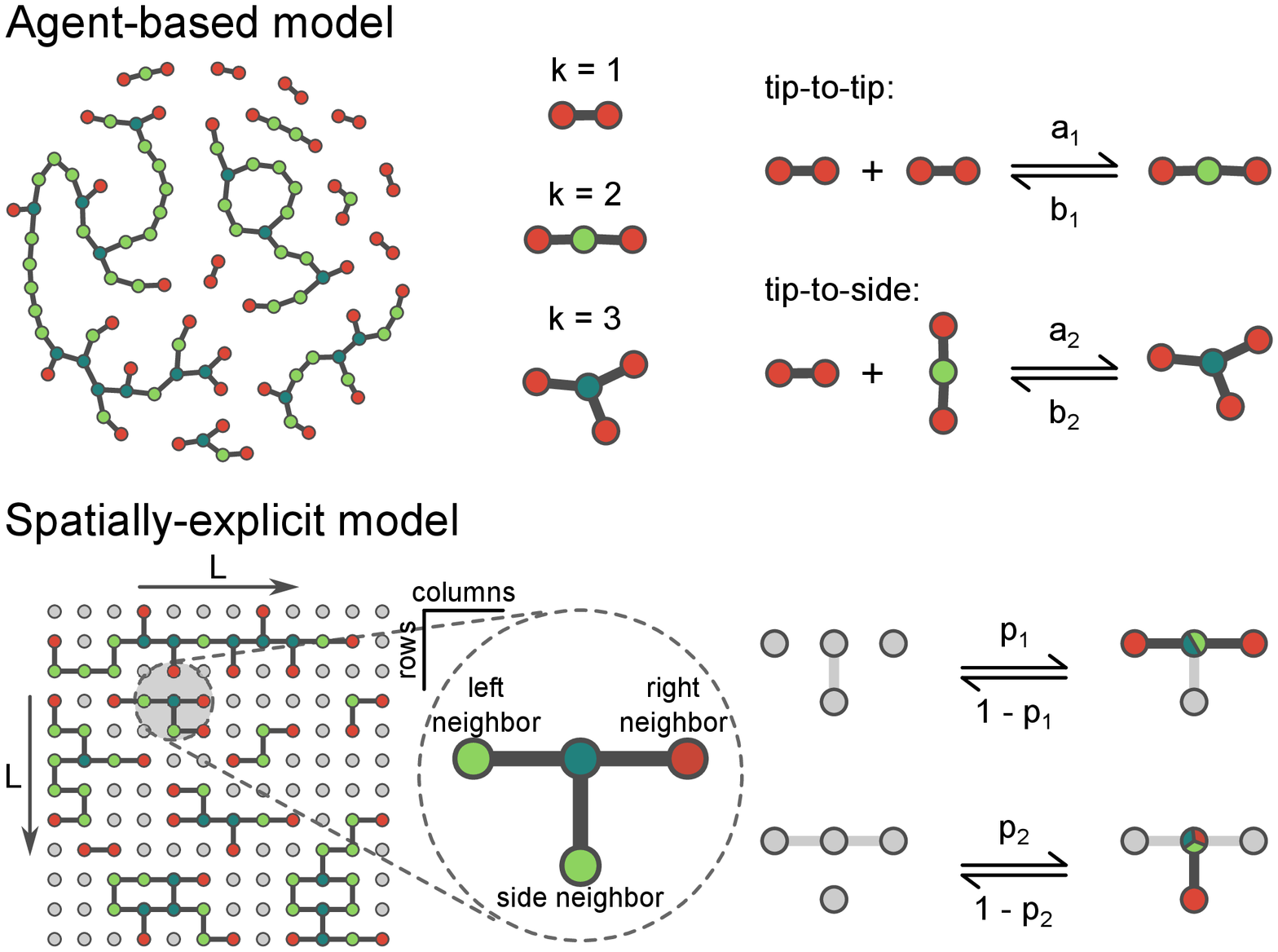}
\caption{The two models of mitochondrial network dynamics used in this work. In the agent-based model (top panels), network nodes do not have explicit spatial coordinates (i. e., it is a dimensionless model). The final topology of the network emerges from the iteration of two types of events: tip-to-tip events, in which two $k=1$ units are merged into a $k=2$ unit (or vice versa) and tip-to-side events, in which a $k=1$ unit and a $k=2$ unit are merged into a $k=3$ unit (or vice versa). In the spatially-explicit model (bottom panels), the network nodes are embedded in a 2-dimensional lattice with predetermined nearest neighborhood interactions. The interactions are anisotropic: a bond is established between a node and its left/right nearest neighbors with probability $p_1$ (or destroyed with probability $1 - p_1$). Similarly, a bond is established between a node and its side nearest neighbor with probability $p_2$ (or destroyed with probability $1 - p_2$).}
\label{models}
\end{figure}

\section{Models definition}
\emph{Agent-based (AB) model:} Here we use a recently introduced agent-based model of the mitochondrial chondriome\cite{sukhorukov} to study the emergence of complex mitochondrial dynamics from the interactions between network edges (Fig.~\ref{models}). The model assumes three types of nodes: free ends of mitochondrial segments ($k = 1$), bulk sites ($k=2$) and branching points ($k=3$). Links between nodes (edges) represent minimal mitochondrial fragments and define the spatial scale of the network.\\
Model dynamics evolve through tip-to-tip and tip-to-side fission/fusion reactions of the type
\begin{eqnarray}
	2\,X_1 & \rightleftharpoons & X_2 \\
	X_1 + X_2 & \rightleftharpoons & X_3		
\end{eqnarray}
\noindent where $X_i$ ($i=1,2,3$) corresponds to nodes with degree $i$. Tip-to-tip reactions happen with association (dissociation) rate $a_1$ ($b_1$) between a random pair of nodes with degree $k = 1$ (association) or a random site with degree $k = 2$ (dissociation). Tip-to-side reactions happen with association (dissociation) rate $a_2$ ($b_2$) between a random pair of nodes with degrees $k = 1$ and $k = 2$ (association) or for a random site with degree $k = 3$ (dissociation). Following Sukhorukov et al.\cite{sukhorukov}, we take into account that only one type of fission is found experimentally \cite{nota1, hoppins} and assume $b_2=(3/2)b_1$ and varied the relative rates $c_i=a_i/b_i$.\\
Despite network edges are the minimal (indivisible) elements of the model, analogous to the smallest mitochondrial fragment found in nature, fusion and fission processes correspond to network node's transformations, analogous to the cellular machinery responsible to fuse and/or excise mitochondrial segments. Notice that the AB model does not include any geographical detail about nodes' positions, and as a consequence, it does not account for the spatial association between mitochondrial clusters.

\emph{Spatially-explicit (SE) model:} To determine if spatial interactions are sufficient to generate the type of complex behavior observed in mitochondrial networks, we derived a simple lattice model inspired in the biological mechanisms determining the spatial relations of mitochondrial fragments during fission and fusion events.
In the SE model, nodes are embedded in a 2D lattice in which their positions are fixed (Fig.~\ref{models}). In contrast to the global nature of the kinetic equations ruling the evolution in the AB model, the SE model evolution is dictated by probabilities concerning the local connectivity of a random selected node. The SE model assumes two types of neighbors (that imply two different types of bonds): the near neighbors, comprising the two nearest nodes within the same lattice row, and the side neighbor, referring to the nearest neighbor within the same lattice column.
At any given time, a bond between a random node and both its left and right neighbors is established with probability $p_1$ (or destroyed with probability $1-p_1$). Similarly, a bond between the same random node and its side neighbor is established with probability $p_2$ (or destroyed with probability $1-p_2$). The fact that the SE model resembles some dynamic version of a classic percolation model is mirrored here as a way to establish a correspondence between the behavior of mitochondrial ensembles and the critical phenomena in percolating systems.

\begin{figure}[!h]
\centering
\includegraphics[width=0.9\textwidth]{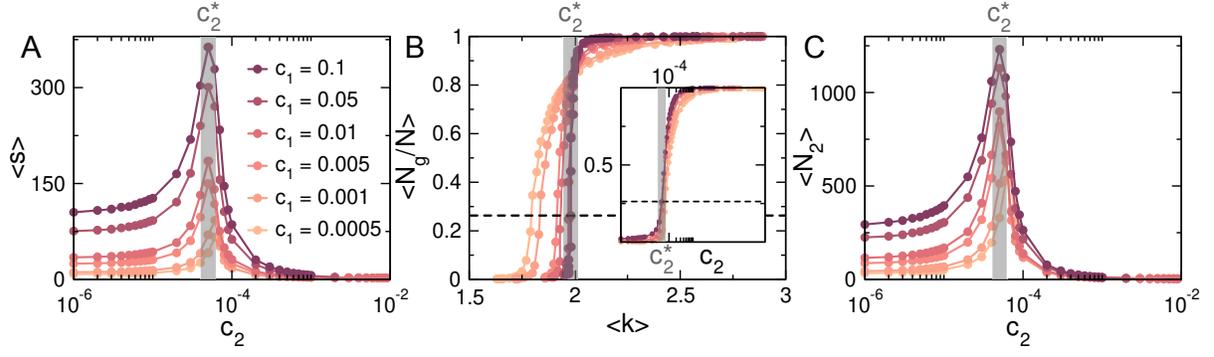}
\caption{{\bf Agent-based model:} Phase transition. A) Average clusters size as a function of $c_2$. B) Order parameter (average fraction of nodes in the largest cluster) {\it vs.} the average degree. Inset shows $\langle N_g/N \rangle$ as a function of $c_2$. Dashed lines denotes the approximate cluster size at the transition). C) Average size of the second largest cluster as a function of $c_2$. Gray regions denote for reference the pseudo--critical threshold. Results obtained from Monte Carlo simulations for $N_e = 15000$ and different values of $c_1$ and $c_2$. }
\label{landscape-ABM}
\end{figure}

\section{Results}
\subsection{Agent-based model}
We first determined the number of iterations required for the AB model to become stationary to be $\approx 2N_e$. We then performed an extensive set of Monte Carlo simulations of the model using the Gillespie algorithm\cite{gillespie}, running every simulation $3N_e$ iterations before measuring the following network quantities: the average degree $\langle k \rangle$ (where the average is taken both over all the nodes in the network and over different runs), the average fraction of nodes in the largest cluster $\langle N_g/N \rangle$ (order parameter of the percolation transition), the average number of nodes in the second largest cluster $\langle N_2 \rangle$ and the average cluster size excluding the largest cluster $\langle s \rangle$ (again the average is taken both over the network and over runs). $\langle s \rangle$ was calculated using the expression from classical percolation theory\cite{barrat}, namely if $N_s$ is the number of clusters of size $s$ and $n_s=N_s/N$, then
\begin{equation}
	\langle s \rangle = \frac{\sum^{'}_s s^2 n_s}{\sum^{'}_s s n_s}
	\label{sus_ms}
\end{equation}
\noindent where the primed sums exclude the largest cluster in the network. Finally, we also computed the complementary cumulative distribution function (CCDF) associated to $n_s$, namely,
\begin{equation}
	N_{c}(s) = \sum^{'}_{s' \ge s} n_{s}(s'),
	\label{ccdf}
\end{equation}
\noindent where the primed sum excludes the giant cluster.

\subsubsection{Phase transition}
We characterized the dynamics of the AB model by studying the behavior of the key network properties near the percolation transition, namely $\langle s \rangle$, $\langle N_g/N \rangle$, and $\langle N_2 \rangle$. Fig.~\ref{landscape-ABM} illustrates the typical behavior of these quantities as a function of the control parameter $c_2$ for different values of $c_1$, keeping the system's size $N_e$ fixed. In agreement with previous results, the average cluster size $\langle s \rangle$ exhibits a maximum at a pseudo--critical value of the control parameter $c_2^*$ (for any value of $c_1$), as shown in Fig.~\ref{landscape-ABM}A.

The phase transition is clearly revealed by the behavior of the order parameter $\langle N_g/N \rangle$, both as a function of $c_2$ (inset) and the mean degree $\langle k \rangle$ (main plot) as shown  in  Fig.~\ref{landscape-ABM}B. Notice the value of $\langle k \rangle < 2$ at the onset of percolation for a four decades range of values of $c_1$.  The size of the second largest cluster $\langle N_2 \rangle$ shows a behavior similar to $\langle s \rangle$, with a peak that accompanies the emergence of the giant cluster, as depicted in Fig.~\ref{landscape-ABM}C.

These results demonstrate the existence of a dynamical phase transition in the AB model, at which the ensemble develops long-range order in the form of a coordinated ``giant'' cluster that, at any given time, involves $\approx 0.3N_e$ connected segments (dashed lines in Fig.~\ref{landscape-ABM}B).

\begin{figure}[h!]
\begin{center}
\includegraphics[scale=.8]{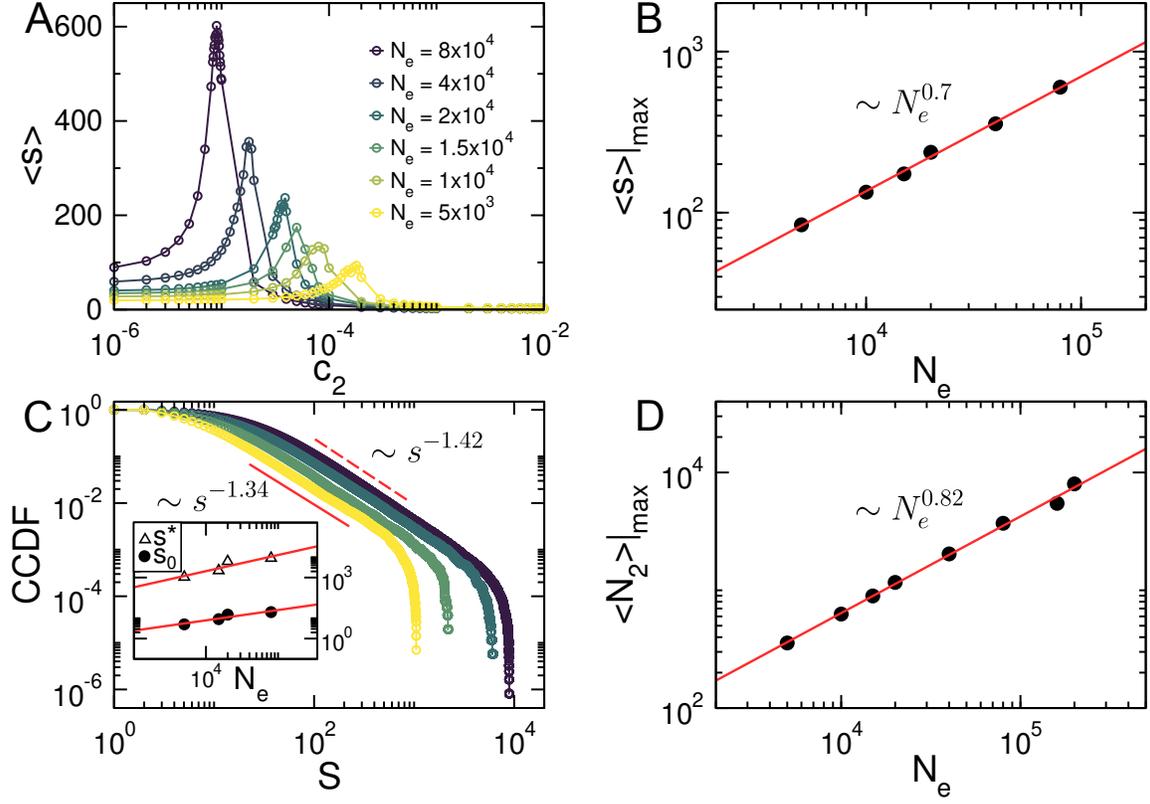}
\caption{{\bf Agent-based model:} Finite size scaling analysis. Monte Carlo simulations for $c_1 = 0.01$ and different system sizes. A) The average clusters size as a function of $c_2$ exhibits a size dependent maximum at a pseudo--percolation threshold $c_2^*(N_e)$. Inset depicts a similar behavior for the susceptibility. B) Log-log plot of the maximum of $\langle s \rangle $ as a function of $N_e$. The straight line corresponds to a power law fitting with exponent $0.7 \pm 0.01$. C) CCDF at $c_2^*(N_e)$ for the same sizes $N_e$ used in panel A. The straight lines are a guide to the eye and correspond to power laws with exponents obtained through a power law fitting of the central part of the CCDF in the two extremes values of $N_e$. The inset shows the scaling with the system size of the low ($s_0$) and large size ($s^*$) cutoffs of the distributions. D) Log-log plot of the average of the second largest cluster as a function of the system size.  The straight line corresponds to a power law fitting with exponent $0.82 \pm 0.01$, in agreement with the exponent of the cutoff $s^*$, since both quantities are to exhibit the same finite size scaling\cite{borgs} $\sim N_e^{d_f/d}$, where $d_f$ is the fractal dimension of the percolating cluster.}
\label{FSS-ABM}
\end{center}
\end{figure}

\subsubsection{Finite-size scaling and universality}
The emergence of a coherent ensemble in the form of a dynamic giant cluster resembles the type of collective behavior characterizing many physical and biological systems in which correlations are amplified in the vicinity of the critical point\cite{cavagna1, attanasi, tang}. However, the critical point is only sharply defined in the thermodynamic limit, away from which, the (effective) critical value of the control parameter depends on the system size. Consequently, quantities like $\langle s \rangle$ and $\langle N_2 \rangle$ are expected to exhibit size-dependent maxima that scale as a function of the system size as $\langle s \rangle|_{max} \sim N_{e}^{\gamma / \nu d}$ and $\langle N_{2} \rangle|_{max}\sim N_{e}^{d_{f}/d}$, respectively. Here, $\gamma$ and $\nu$ are the standard susceptibility and the correlation length critical exponents, respectively, $d$ is the effective dimension of the system (equal to the spatial dimension $D$ if $D < d_{c}$ (the upper critical dimension) or to $d_c$, otherwise) and $d_f$ is the fractal dimension of the percolating cluster. The specific values of these parameters determine the universality class to which the system under scrutiny belongs.

To determine the the critical exponents of the AB model, we performed a finite-size scaling analysis of the network quantities already described. As shown in Fig.~\ref{FSS-ABM}A, for different values of $N_e$, $\langle s \rangle$ exhibits a series of maxima $\langle s \rangle|_{max}$ at pseudo--critical values of the control parameter $c_2^*$. As $N_e$ increases, the peaks become sharper and $c_2^*$ decreases as $c_2^* \sim 1/N_e$ (not shown).
As depicted in Fig.~\ref{FSS-ABM}B the magnitude of $\langle s \rangle |_{max}$ (i. e., its value at $c_2 = c_2^*$) follows a power law with exponent $\gamma / \nu d \sim 0.7 \pm 0.01$.
To further understand the nature of the transition, we studied the behavior of the cluster size distribution at the critical point. Fig.~\ref{FSS-ABM}C shows the CCDF at the pseudo--percolation threshold for different system sizes.
At variance with the expectations for a  percolation based model namely, a power law with an exponential cutoff for large sizes, the CCDF for the AB model is consistent with cluster size distribution with {\em two} cutoffs
\begin{equation}
    n(s) \sim \theta(s-s_0)\, s^{-\tau} \ e^{-s/s^*},
\end{equation}
\noindent where $\tau$ corresponds to Fisher exponent, the term $e^{-s/s^*}$ corresponds to the exponential cutoff with $s^* \propto N_2$ and $s_0$ is a low size  cutoff (here $\theta(x)$ is the Heaviside step function, namely $\theta(x) = 1$ if $x>0$ and $\theta(x) = 0$ otherwise).  We see that, for intermediate an large scales,  $n(s)$ develops a power law behavior with exponent $\tau = 2.38 \pm 0.04$ ($N_c \sim s^{-(\tau-1)}$) followed by a cutoff at size $s^*(N_e)$. Both cutoff $s^*$ and $s_0$ scale as a power law with the system size  with  exponents $0.80 \pm 0.2$ and $0.5 \pm 0.1$ respectively. Hence, the presence of the small size cutoff is a strong finite size effect which disappears for large enough system sizes. However, this has a strong impact on the estimation of the critical exponents at the scales of interest for mitochondria. For instance, for the feasible simulation sizes,  the estimated values of  $\tau \approx 2.4$ and the cutoff exponent $d_f/d=0.80 \pm 0.2$ are consistent with the standard percolation mean field values\cite{stauffer} $\tau=5/2=2.5$ and $d_f/d = 2/3 \approx 0.67$, an expected result in a dimensionless system. However, the estimated value of the exponent $\gamma/\nu d=0.7 \pm 0.1$ (see Fig.~\ref{FSS-ABM}B) is not consistent with the mean field value $\gamma/\nu d=1/3$. This is a direct consequence of the above mentioned finite size effect.

\begin{figure} [h!]
\centering
\includegraphics[width=1\textwidth]{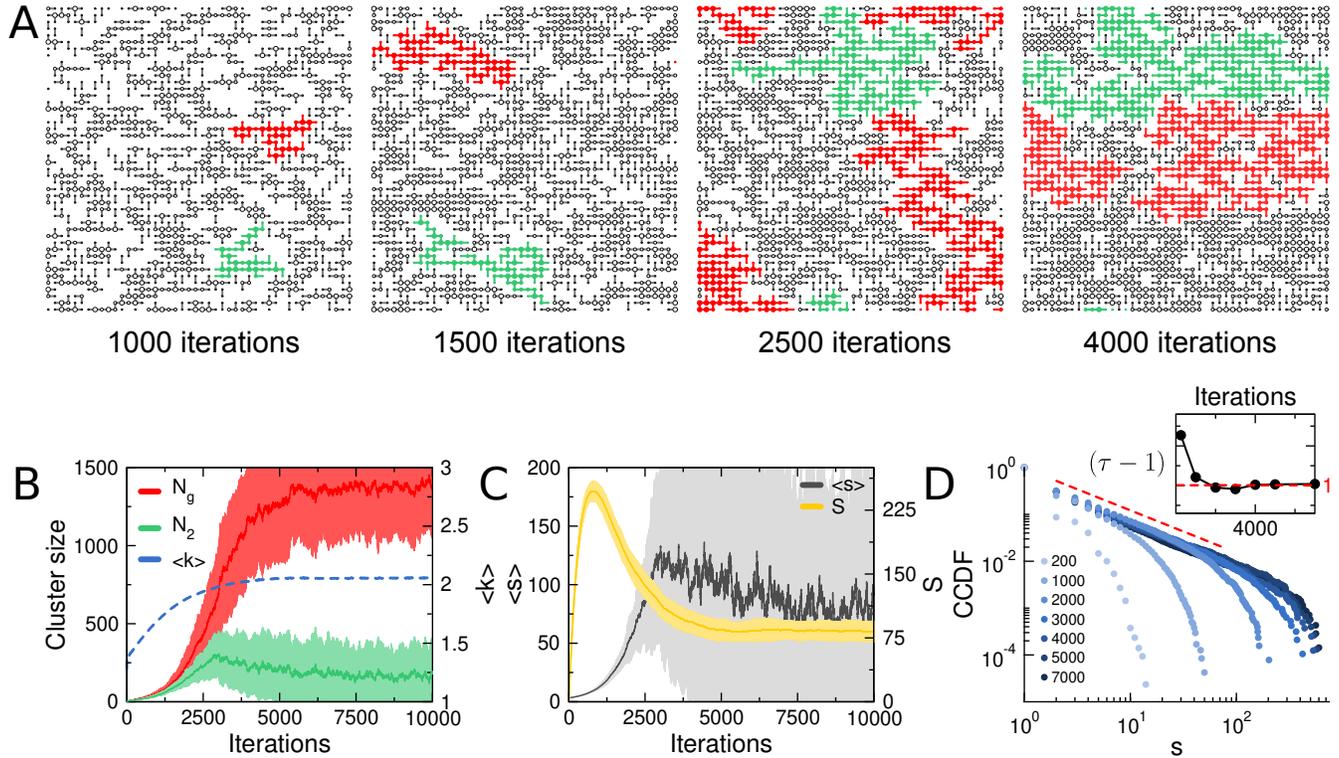}
\caption{{\bf Spatially-explicit model:} Temporal evolution of network quantities. Monte Carlo simulations for $N = 2116$, $p_1 = 0.65$ and $p_2 = 0.5$. A) Snapshots illustrating the emergence of a ``giant'' cluster (shown in blue). B) Evolution of the size of the ``giant'' cluster, the size of the second largest cluster and the average degree as a function of the number of iterations. After $\approx 1.5N$ iterations the system undergoes a phase transition characterized by the emergence of a stable ``giant'' cluster. The average degree stabilizes at $\approx 2$. C) Temporal evolution of the average cluster size and the number of clusters. After the transition the number of clusters decreases monotonically until it reaches a stable value. The average cluster size, on the contrary, exhibits large fluctuations illustrating the fact that after the transition the system is poised at a highly susceptible state. D) Evolution of the cluster distribution as a function of the number of iterations. Inset shows how the exponent of the distribution converges to $\approx 1$ (which corresponds to $\tau \approx 2$).}
\label{SEM}
\end{figure}

\subsection{Spatially-explicit model}
We implemented the SE model on a square lattice of size $N = L^2$ in which links between sites are constantly being created and/or destroyed obeying the following rules (Fig.~\ref{models}B): {\it i}) links between the $i$-est node and both its left and right neighbors are established (or destroyed) with probability $p_1$ ($1 - p_1$) and {\it ii}) a link between the $i$-est node and its side neighbor is established (or destroyed) with probability $p_2$ ($1 - p_2$).\\
The snapshots in Fig.~\ref{SEM}A illustrate the topological evolution of the SE model for $N = 2116$, $p_1 = 0.65$ and $p_2 = 0.5$. Fig.~\ref{SEM}B depicts the size of biggest cluster $\langle N_g \rangle$, the size of the second biggest cluster $\langle N_2 \rangle$ and the mean degree $\langle k \rangle$ as a function of the number of iterations. After an initial growth of $\langle N_g \rangle$ and $\langle N_2 \rangle$, the magnitude of both these quantities become stationary after $\approx 2.5N$ iterations (dashed line). $\langle k \rangle$ converges to a value $\approx 2$ approximately at the same time.\\
We then studied the behavior of the network clusters by computing both the total number of clusters $S$ and the mean cluster size $\langle s \rangle$ as a function of the number of iterations. As shown in Fig.~\ref{SEM}C, $S$ initially grows rapidly, as small clusters emerge everywhere in the network, and later decreases, as growing clusters start to coalesce, slowly converging to a fixed value. On the other hand, $\langle s \rangle$ (computed using Eq.~\ref{sus_ms}) builds up for $\approx N$ iterations to later display large fluctuations. This behavior is expected since these fluctuations grows towards criticality and they are maximal at the critical point\cite{tang, stauffer}.\\
Finally, we corroborated that the cluster size distribution converges to a power law (up to a large scale cutoff) with the number of iterations, as shown in Fig.~\ref{SEM}D. As depicted in the inset, the exponent of the distribution converges after $\approx 2N$ iterations.

\begin{figure} [h!]
\centering
\includegraphics[width=0.9\textwidth]{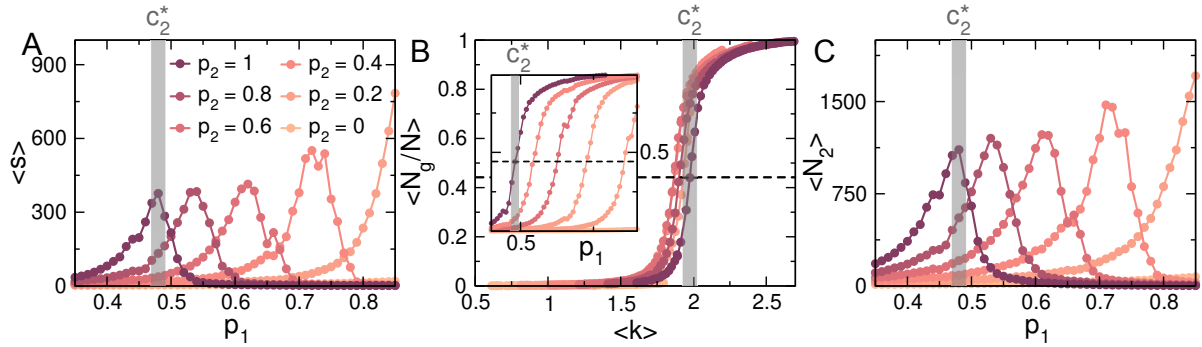}
\caption{{\bf Spatially-explicit model:} Phase transition.  A) Average clusters size as a function of $p_1$. B) Order parameter (average fraction of nodes in the largest cluster) {\it vs.} the average degree. Inset shows $\langle N_g/N \rangle$ as a function of $p_1$. C) Average size of the second largest cluster as a function of $p_1$.  Results computed from Monte Carlo simulations for $N = 2116$ and different values of $p_2$.}
\label{landscape_SEM}
\end{figure}

\subsubsection{Phase transition}
We now study the behavior of the SE model near the percolation transition. Fig.~\ref{landscape_SEM} illustrates the typical behavior of the quantities $\langle s \rangle$, $\langle N_g/N \rangle$ and $\langle N_2 \rangle$ as a function of $p_1$, for different values of $p_2$. It is clear that, differently from the AB model, the control parameter in the SEM is $p_1$ (roughly equivalent to $c_1$ in the AB model). Given a value of $p_2$, $p_1$ controls the percolation transition, characterized as a peak in $\langle s \rangle$, as shown in Fig.~\ref{landscape_SEM}A (gray region is centered at the pseudo--critical threshold for $c_2 = 1$ as reference), which happens when the order parameter $\langle N_g/N \rangle$, depicted in Fig.~\ref{landscape_SEM}B, is $\approx 0.6$.
Fig.~\ref{landscape_SEM}C shows the average size of the second largest cluster $\langle N_2 \rangle$ as a function of $p_1$. Notice that $p_{1}^{*}$ increases as a function of $p_2$, matching the behavior of $\langle N_g \rangle$ (inset in Fig.~\ref{landscape_SEM}B).

\begin{figure} [hb!]
\centering
\includegraphics[width=.7\textwidth]{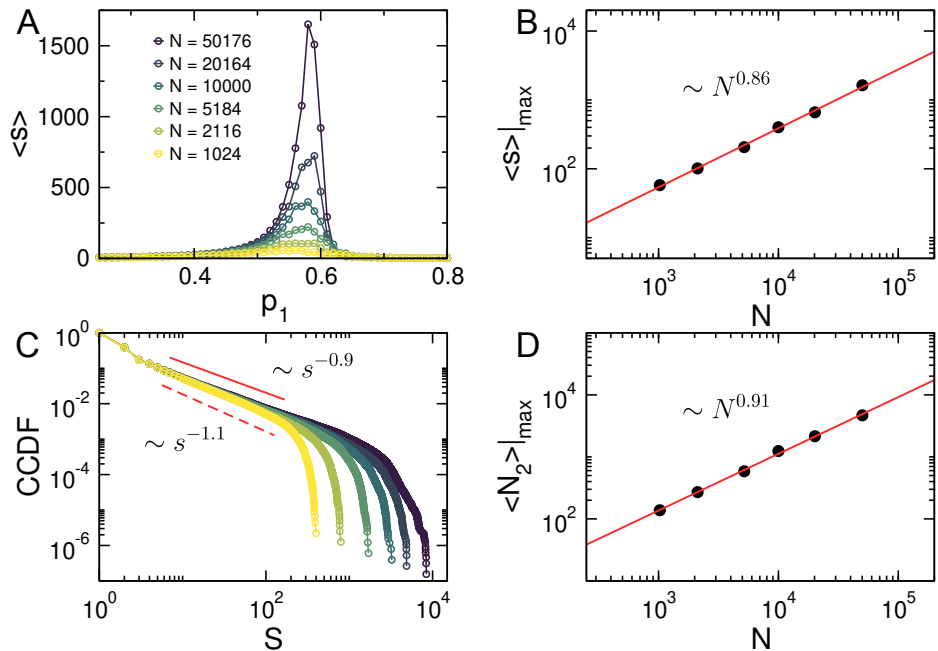}
\caption{{\bf Spatially-explicit model:} Finite size scaling behavior of the relevant quantities for $p_2 = 0.7$. A) Average clusters size $\langle s \rangle $ as a function of $p_1$ for different system sizes. The curves exhibit a size dependent maximum at a pseudo--percolation threshold $p_1^*(N)$ B Log-log plot of the maximum of $\langle s \rangle $ as a function of $N$. The straight line corresponds to a power law fitting. C) CCDF at $p_1^*(N)$ for the same sizes $N$ used in panel A. The straight lines are a guide to the eye and correspond to power laws with exponents obtained through a power law fitting of the central part of the CCDF in the two extremes values of $N$. D) Log-log plot of the average of the second largest cluster as a function of the system size. The straight line corresponds to a power law fitting. }
\label{FSS-SEM}
\end{figure}

\subsubsection{Finite size scaling and universality}
We repeated for the SE model the same finite size scaling analysis we performed for the AB one. In Fig.~\ref{FSS-SEM} we show the behavior of the different quantities, as we vary the system size $N$. As in the case of the AB model, we observed the expected  scaling behaviors of the maxima of $\langle s \rangle$ and $N_2$. In this case, the estimations of the corresponding critical  exponents were $\gamma / \nu d \approx 0.86 \pm 0.1$ and $d_f/d \approx 0.91 \pm 0.1$.
In Fig.~\ref{FSS-SEM}C, the behavior of the CCDF at the pseudo--percolation threshold $p_1^*$ as a function of $N$ is shown. In this case the cluster sizes distribution exhibit the expected power law behavior with only one cutoff at large cluster sizes $s_* \propto N_2$

\begin{equation}
    N_c(s) \sim  s^{-(\tau-1)} \ e^{-s/s^*},
\end{equation}

\noindent where $\tau = 2.0 \pm 0.1$. We see that the whole set of exponents is fully consistent  with the universality class of standard 2D percolation: $\tau=187/91\approx 2.055$, $\gamma / \nu d =43/48\approx 0.896$ and   $d_f/d =91/96 \approx 0.948$.

\subsection{Mitochondrial networks from mouse embryonic fibroblasts}
We now turn our attention to real mitochondrial networks, from mouse embryonic fibroblasts (MEFs). Our hypothesis is that the topology of these networks corresponds to network configurations at criticality. If that is the case, networks of different masses should correspond to the peaks in $\langle s \rangle$ shown in Figs.~\ref{FSS-ABM}A and \ref{FSS-SEM}A. Consequently, our prediction is that mitochondrial networks of increasing masses will show higher susceptibility $\langle s \rangle$ and more massive second biggest clusters $N_2$. To test this, we took advantage of the fact that mitochondrial mass varies from cell to cell, and we quantified the finite-size effects in several quantities from images of mitochondrial networks from MEFs using a modified version of the procedure described in \cite{zamponi} (see Methods).

\begin{figure} [h!]
\centering
\includegraphics[width=.7\textwidth]{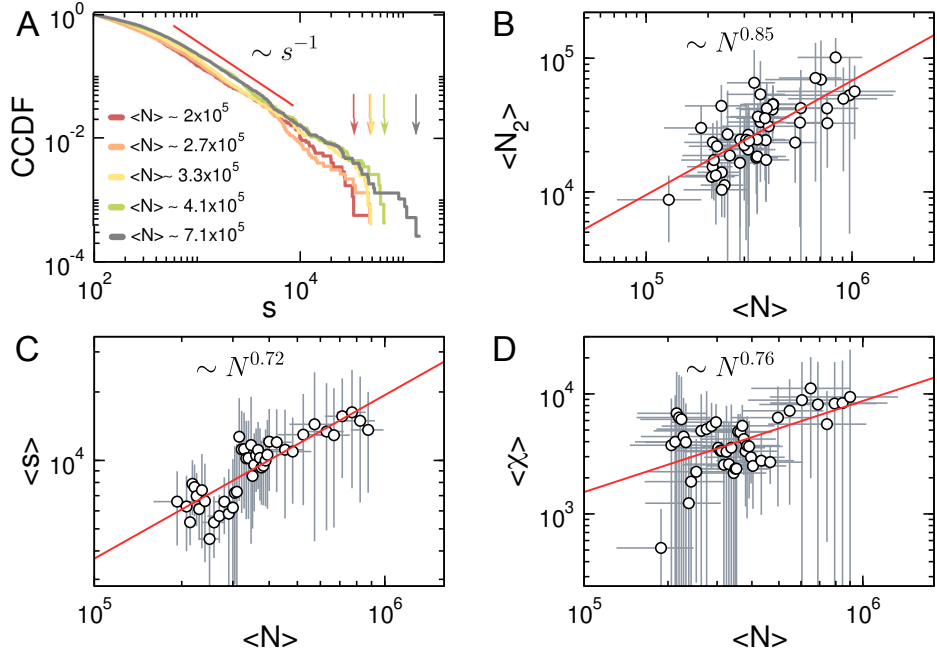}
\caption{{\bf Mouse embryonic fibroblasts mitochondrial networks:} Finite-size scaling  A) The complementary cumulative distribution function (CCDF) of the mitochondrial cluster mass obeys a power-law with exponent $\tau \approx 2$, independently of the average total mass of the network $\langle N \rangle$. Finite size effects are evidenced in the different cutoff values $s^*$ at the tail of the distributions (arrows). B) The size of the second largest cluster $\langle N_2 \rangle$ scales with $\langle N \rangle$ as $\langle N_2 \rangle \sim \langle N \rangle^{\omega_1}$ with $\omega_1 \approx 0.85 \pm 0.11$. C) The susceptibility $\langle s \rangle$ (Eq.~\ref{sus_ms}) as a function of $\langle N \rangle$ follows a power-law with exponent $\omega_2 \approx 0.72 \pm 0.1$. $\langle N \rangle$ corresponds to the average total mitochondrial mass estimated from images, utilized here as a proxy for network size. Symbols correspond to mean values and error bars to standard deviations from different thresholds. D) The alternative estimation of the susceptibility $\chi$ (Eq.~\ref{sus_chi}), as a function of $\langle N \rangle$ follows a power-law with exponent $\omega_2 \approx 0.76 \pm 0.18$.}
\label{FSS-exp}
\end{figure}

Using the definition in Eq.~\ref{ccdf} we computed the CCDF of the cluster mass, which is equivalent to the cluster size  in the models. We proceeded by sorting first all network configurations based on their total pixel mass and then to divide the set in 5 equal subsets comprising 10 network configurations each. Being each network a cluster distribution in itself, giant clusters were excluded before the CCDF computation (see Eq.~\ref{ccdf}). The distributions corresponding to subsets are shown in Fig.~\ref{FSS-exp}A, from which two different regimes can be distinguished: {\it i}) a power law regime that spans for almost two decades with exponent $\tau \approx 2$; and {\it ii}) a mass-dependent exponential cutoff.
At criticality, the cutoff $s^*$ and the maximum of $\langle N_2 \rangle$ are expected to scale as a power law with the system's size  $N^{\omega_1}$. To test if this is the case, we applied a sliding window of size $n$ along the sorted network configurations and calculated the average mass of the second biggest cluster for each set of $n$ networks. Results are presented in Fig.~\ref{FSS-exp}B, from which it is possible to conclude that $\langle N_2 \rangle \sim N^{\omega_1}$, with $\omega_1 \approx 0.85 \pm 0.11$. This exponent plays the role  $d_f/d$ in the models, thus allowing for a comparison.

To further validate these findings, we computed $\langle s \rangle$ (Eq.~\ref{sus_ms}), using the same procedure described before. As stated in previous sections, the system's susceptibility at criticality is expected to scale with a power law $\langle s \rangle \sim N^{\omega_2}$. Our results, presented in Fig.~\ref{FSS-exp}C, demonstrate that this is the case in real mitochondrial networks, with $\omega_2 \approx 0.72 \pm 0.11$. As in the case of $N_2$, this exponent should be compared with $\gamma/\nu d$  for the models. In addition to Eq.~\ref{sus_ms}, the system's susceptibility can be computed as the average magnitude of the fluctuations in the order parameter $\langle N_g \rangle$\cite{stauffer} as
\begin{equation}
	\chi = \langle N \rangle \left[\bigg\langle\bigg(\frac{N_g}{N}\bigg)^2\bigg\rangle - \bigg\langle\frac{N_g}{N}\bigg\rangle^2 \right].
	\label{sus_chi}
\end{equation}
As shown in Fig.~\ref{FSS-exp}D, we corroborate that this quantity scales in the same way as $\langle s \rangle$, since we obtain $\langle \chi \rangle \sim N^{\omega_2}$, with $\omega_2 \approx 0.76 \pm 0.18$.\\
Overall, the estimation of scaling exponents computed from the experimental data, to be compared with the theoretical ones, are: $\tau \approx 2 \pm 0.1$, $\gamma / \nu d \approx 0.72\pm0.1$ and $d_f/d \approx 0.85\pm0.1$.

\section{Discussion}
A growing body of evidence supports the hypothesis that biological systems are poised near criticality. That is to say, at a region in the control parameters space characterized by long-range correlations and large susceptibility\cite{mora, munioz}. Such description fits well with the general idea that biological systems need to actively maintain a number of internal variables within a narrow window, between order and disorder, to sustain life\cite{chialvo_acta_polB}. Albeit the attractiveness of this proposal, concrete proof that a certain biological system is poised at criticality is hard to provide, given the general impossibility to both tune the parameter that control the transition and change the size of the system\cite{attanasi, munioz, box}.

In this work, we have approached this issue by first modeling mitochondrial dynamics to find the theoretical boundaries for the universal behavior of mitochondrial networks. Such universality is determined by a set of exponents that describe the behavior of the system independently of its fine-grain details. The exponents are obtained by studying how the relevant quantities of the system vary with the system size (i. e., the scaling relations). To that, we studied two conceptually different models, namely a recently published agent-based model that is capable of generating network topologies that resemble real mitochondrial networks\cite{sukhorukov, zamponi} and a simple spatially explicit model that can be roughly defined as an anysotropic version of the dynamic  percolation model. Then, we took advantage of the `default' mitochondrial mass fluctuations observed in MEFs to measure relevant quantities as a function of different masses as a proxy for a finite-size scaling analysis.

Our results, summarized in Table~\ref{tab:table1}, indicate that the AB model, besides some strong finite size deviations, belongs to the mean-field universality class, while the SE model does so to the standard 2D percolation universality class. Interestingly, though the AB model generates more `realistic' topologies compared to the SE model, the exponents obtained from real networks suggest that they belong to the 2D standard percolation universality class. This suggests  a 2D description of mitochondria to be  more appropriated than dimensionless one. Moreover, the SE model provides a more accurate description of the cluster size distribution than the AB one, due to the above mentioned finite size effects, which cannot be neglected for the typical mitochondrial scale.

\begin{table}
\begin{center}
\begin{tabular}{ |c|c|c|c|}
 \hline
   & $\tau$ & ${\gamma / \nu d}$ & ${d_{f}/d}$\\
   \hline
  2D Percolation & $187/91\approx 2.055$ & $43/48 \approx 0.896$ & $91/96 \approx 0.948$ \\
  Mean field Perc.& $5/2=2.5$ & $1/3 \approx 0.33..$ & $2/3 \approx 0.66..$ \\
 AB model & $2.38\pm0.04$ & $0.7\pm0.01$& $0.82\pm0.01$\\
 SE model & $2\pm0.1$& $0.86\pm 0.1$& $0.91\pm0.1$ \\
 MEF & $2\pm0.1$ & $0.72\pm0.1$ & $0.85\pm0.11$ \\
 \hline
 \end{tabular}
 \caption{\label{tab:table1} Main scaling exponents calculated from the two models and from mitochondria images from MEFs.}
 \end{center}
\end{table}

\section{Methods}
\subsection*{AB model}
The agent-based (AB) model follows closely the implementation described in \cite{sukhorukov, zamponi}.

\subsection*{SE model}
We implemented the SE model on a square lattice of size $N = L^2$. With certain probability, two types of links are established: what we call ``left/right links'' and ``side links''. In a two coordinates system, the right and left nearest neighbors of the $i$-est node (with coordinates $(i, j)$) would be nodes at positions $(i+1, j)$ and $(i-1, j)$, respectively (Fig.~\ref{models}, lower panel). Similarly, the side neighbor of the $i$-est node is the node located at position $(i, j\pm1)$. Notice that these definitions of left/right and side links are made only out of numerical and algorithmic convenience. Two parameters specify how links are established independently: $p_1$ is the probability for a node to be linked with both its left and right neighbors and $1 - p_1$ is the probability for the links between the $i$-est node and both its left and right neighbors to be destroyed (analogous to tip-to-side reactions in the AB model). In parallel, $p_2$ is the probability for a link between the $i$-est node and its side neighbor to be created and $1 - p_2$ is the probability for the link between the $i$-est node and its side neighbor to be destroyed (analogous to tip-to-side reactions in the AB model).

\subsection*{Cell culture}
The maintenance and procedures of all animals were in accordance with and approved by the Research Animal Resource Center of the Weill Cornell Medical College. Mice used for experiments were between 8-14 weeks of age. Mouse embryonic fibroblasts (MEFs) were obtained as previously described \cite{xu}. Briefly, 13.5 days pregnant female C57BL6 mice were sacrificed by CO$_{2}$ inhalation followed by cervical dislocation. Uterus was removed and embryos harvested, rinsed with PBS and placed on a petri dish. The head and red organs were discarded and the remaining tissue was chopped with razor blades and trypsinized for 15 min at $37{^\circ}$C. Trypsin reaction was quenched with Dulbecco's Modified Eagle Medium (DMEM - GIBCO) supplemented with 10\% FBS (GIBCO) and the mixture was centrifuged for 5 min at 2000 rpm. The cellular pellet was resuspended in DMEM supplemented with 10\% FBS, 1\% L-Glutamine (Sigma Aldrich), 1\% Sodium Pyruvate (Sigma Aldrich), 1\% HEPES (Sigma Aldrich) and 1\% Penicillin/Streptomycin (GIBCO). Cell pellets from 4 embryos were seeded on 175 cm$^{2}$ culture bottles and allowed to grow for 48 h in a 5\% CO$_2$ and $37{^\circ}$C atmosphere.\\
Human Embryonic Kidney (HEK) 293T cells were cultured in DMEM  supplemented with 10\% FBS, 1\% L-Glutamine, 1\% Sodium Pyruvate, 1\% HEPES and 1\% Penicillin/Streptomycin in a 5\% CO$_2$ and $37{^\circ}$C atmosphere.

\subsection*{Lentivirus generation and infection}
The mitochondrial-targeted YFP plasmid was purchased from OriGene. The ORF containing both the mitochondrial targeting sequence and the YFP was subcloned into the lentiviral plasmid pLV-eGFP (Addgene \#36083) to shield pLV-mitoYFP.\\
Lentiviral particles carrying the pLV-mitoYFP construct were produced as described before \cite{baloh}. Briefly, HEK 293T cells were plated onto six-well plates and transfected with a polymerase-coding vector (REV), a packaging vector (8.71), an envelope vector (VSVG) and the shuttle vector pLV-mitoYFP using Lipofectamine 2000 (Invitrogen). Media was changed at 12 h and collected at 48 h and 72 h, pooled and applied directly to MEFs cultures.

\subsection*{Imaging}
Images were collected on a Zeiss LSM 880 microscope equipped with the AiryScan detector using a 63x/1.4 NA Plan-Apochromat Oil DIC M27 objective lens (Zeiss) and an Edge 5.5 sCMOS camera (PCO). YFP was excited with a 488 nm Argon laser (5\% power, $\sim 60 \mu W$) and collected with a 500/550 nm emission filter. Gain was set to 800. Scan mode was set to Frame with optimal frame size (3812x3812 pixels) resulting in an image pixel size of $\sim 35.29$ nm and a lateral resolution of $\sim140$ nm. Speed was set to 8 ($2\mu s$ Pixel Dwell time) and the Bit depth at 16 bits. Prior to image analysis, raw `.czi' files were processed into deconvoluted Airyscan images using the Zen software with default settings.

\subsection*{Image analysis}
Data extraction from images was performed using a custom-written MATLAB code that extracts network quantities from `.czi' files produced by the ZEN software. The script first converts raw images to binary data by performing image thresholding. Subsequently, individual clusters are identified as groups of pixels connected by at least one of the eight nearest neighbors. These procedure yields a cluster distribution for each network analyzed from which all the relevant quantities used in this study can be obtained. Note that in this case the skeletonization step from \cite{zamponi} was skipped to better estimate the mass of each cluster.

\subsection*{Code availability}
Codes to implement both models and a script to reproduce Fig.~\ref{FSS-exp} is available at \textcolor{blue}{https://github.com/nahuelzamponi/mtmodels}


\begin{thebibliography}{9000}
\bibitem{nota1} To be more precise, though the current description of mitochondrial fission does involve different protein complexes \cite{hoppins}, resulting fission events are indistinguishable from each other, showing no particular specificity for linear fragments ($k=2$) or branching points ($k=3$).
\end{thebibliography}
\end{document}